# Petri Net Modeling of the Brain Circuit Involved in Aggressive Behavior

# Modelo descriptivo del circuito cerebral involucrado en la agresión mediante redes de Petri

Johann Heinz Martínez Huartos

Facultad de Administración, GIPE, NEUROS, Universidad del Rosario. Bogotá Colombia.

Dirección Correspondencia: Dpt. Ciencia y Tecnología Aplicadas a la I.T.A. E.U.I.T. Agrícola, UPM Ciudad Universitaria s/n 28040 Madrid, España. Tel: 91 336 3720

johann.martinez@urosario.edu.co

Carlos B. Moreno

Escuela de Medicina y Ciencias de la Salud, Grupo NEUROS Universidad del Rosario. Bogotá Colombia.

carlos.moreno@urosario.edu.co

José Luis Díaz

Departamento de Historia y Filosofía de la Medicina, Facultad de Medicina, Universidad Autónoma de México. México, D.F., México. jldiaz43@gmail.com

**Resumen**

El propósito de este trabajo es mostrar los resultados iniciales de una línea de investigación que tiene como objetivo desarrollar modelos dinámicos del circuito cerebral implicado en la conducta agresiva y sus estructuras asociadas. De esta forma, se pretende esquematizar por medio de redes de Petri (una herramienta computacional tipo "work flow") el complejo proceso nervioso usualmente correlacionado con emociones de rabia y que desemboca en comportamiento agresivo. En primer término se presenta la técnica de modelamiento tomando en cuenta la noción más actual y aceptada del substrato neurobiológico de las emociones, en especial de las estructuras involucradas en el circuito de la agresión, sus entradas, salidas y conectividad. A continuación se muestran los fundamentos operativos para la posible simulación de sistemas dinámicos basada en redes de Petri que aquí se utilizan para modelar esas estructuras, sus conexiones y su dinámica en el tiempo. Finalmente, se presenta el modelo resultante y la simulación del proceso nervioso asociado con la conducta agresiva, lo cual se propone que otorga un sustento experimental *in silico* a la Teoría de los Procesos Pautados.

## Summary

The purpose of this work in to demonstrate the initial results of a research project having as its goal to develop dynamic models of the brain network involved in aggressive behavior. In this way, the complex neural process correlated to basic anger emotions and resulting in aggressive behaviors is purportedly schematized by the use of Petri nets, a work-flow computational tool. Initially, the modeling technique is introduced taking into account the most recent and accepted notion of the neural substrates of emotions, particularly the brain structures involved in the aggression neural network, including their inputs, outputs, and internal connectivity. In order to optimally represent these structures, their connections, and temporal dynamics, the Petri net foundations employed in the simulation theory are defined. Finally, the model and dynamic simulation of the neural process associated with aggressive behavior is presented and evaluated as a feasible *in silico* experimental support of the Patterned-Process Theory.

# Introducción

El término agresión designa a un conjunto de conductas nocivas dirigidas por individuos usualmente incitados por la emoción básica de enojo, rabia o furia hacia uno o varios receptores que amenazan con producir o de hecho provocan temor, dolor y lesión. Son conductas interactivas y sociales que en contextos agonistas de confrontación se manifiestan como amenazas, embates, persecuciones, prendimientos, golpes, laceraciones o mordidas y que pueden ser respondidos de formas tan diversas como defensa, contraataque, pelea, lamento, protesta, sumisión, derrota, evasión o huida [1]. Dado su destacado papel en la territorialidad, la selección y competencia sexual, la dominancia, o el control de los recursos, la agresión es una conducta relevante en el proceso evolutivo de la mayoría de las especies animales y su desarrollo ha contribuido a la adaptación y preservación de la especie humana [2,3]. La agresión se ha estudiado ampliamente en diversas disciplinas, desde aproximaciones etológicas tanto cualitativas como cuantitativas en diferentes especies [3,4], hasta estudios anatómicos y funcionales de las estructuras cerebrales implicadas en su expresión [5,6,7,8,9,10]. El hallazgo y modelaje de los sistemas cerebrales involucrados en este tipo de comportamientos ha constituido una ruta de investigación creciente y promisoria para comprender mejor las bases cerebrales de la conducta [10, 11, 12, 13, 14, 15, 16, 17, 18].

Entre las propuestas actuales para abordar la generación de los correlatos cognitivos y cerebrales del comportamiento está la teoría de los procesos pautados [19]. Esta teoría permite modelar el correlato cerebral de la agresión pues propone herramientas epistemológicas y metodológicas para establecer una relación funcional y dinámica entre los procesos cerebrales, conductuales y mentales. Según la teoría, estos tres fenómenos comparten una morfología dinámica similar que permite modelarlos utilizando herramientas matemáticas de los sistemas complejos. Esto es así porque un proceso pautado implica una transición de estados en el espacio y en el tiempo que transporta y transforma información relevante mediante un desarrollo funcional entre las estructuras o condiciones involucradas. La identificación de los elementos constitutivos del sistema, el cambio de estados y el transporte de información en un sistema abierto de tipo red permite modelarlos con herramientas funcionales y representarlos en diagramas dinámicos. En el caso del sistema nervioso, las pautas inter-modulares de actividad cursarían entre los módulos cerebrales de una forma semi-ordenada y se acoplarían de una manera lo suficientemente compleja como para que ocurra un correlato mental.

La herramienta propuesta inicialmente para modelar los procesos pautados se denomina redes de Petri, grafos bi-partitos que sirven para hacer análisis tanto empíricos como teóricos de sistemas complejos concurrentes [20]. Las redes de Petri tienen dos tipos de nodos, llamados *lugares* y *transiciones*. Los lugares o plazas P, se representan como círculos separados por una transición T, que se representa por líneas rectas cortas o rectángulos; los lugares se unen a las transiciones por medios de vectores o *Arcos*. Este tipo de gráficos tienen además la capacidad de almacenar en sus lugares *fichas* o *tokens* K, marcadas como puntos rellenos que constituyen la parte dinámica que simula las actividades concurrentes en un sistema dinámico **(Figura 1)**[21]. La dinámica temporal de la red se procesa mediante "disparos" por medio de los cuales los tokens ubicados en una plaza o nodo se transportan a otra plaza a través de una transición marcada entre ellas. Este comportamiento dinámico es particularmente relevante y ajustable a un circuito nervioso en el cual la información pasa de un módulo a otro mediante las vías aferentes y eferentes de conexión entre los módulos involucrados de tal forma que, los tokens transferidos de una plaza a otra representan a la información nerviosa transmitida entre los módulos representados por las plazas y al mismo tiempo el estado de activación de cada plaza entre los disparos de la red.

Las redes de Petri se usan para visualizar procesos de distribución de información y comunicación [22]. Además, es posible obtener ecuaciones que gobiernen transiciones de estado de un sistema [23, 24]. Las reglas para la dinámica de las redes de Petri se resumen en tres: (1) Una transición t está habilitada si todos los lugares de entrada tienen al menos un token; (2) una transición habilitada se puede disparar moviendo un token de cada lugar de entrada y poniéndolo en algún lugar de salida y (3) cada disparo genera un nuevo estado y por ende un cambio en el vector marcador o de estados del sistema en una red.

A partir de estos desarrollos, de los modelos de sistemas dinámicos y de la teoría de procesos pautados, se pretende en el presente trabajo modelar la dinámica del sistema de núcleos cerebrales involucrados en la conducta agresiva utilizando redes de Petri. El trabajo es parte de un proyecto que tiene como objetivo realizar una simulación computacional que aplique la teoría de los procesos pautados a sistemas cerebrales que tienen correlatos tanto conductuales y expresivos como cognitivo-emocionales.

# Materiales y métodos

## Correlato nervioso de la agresión y modelo funcional

Para especificar el sistema cerebral involucrado en la génesis de la conducta agresiva se utilizó el modelo propuesto por Nelson y Trainor [25] **(Figura 2a)**. En este modelo la actividad del circuito se origina en la corteza cerebral en respuesta a diversos estímulos del medio exterior que llegan a las zonas sensoriales primarias. Estas zonas sirven a su vez como eferentes de una información que se bifurca para alcanzar a la corteza órbito-frontal (OFC) y al núcleo medial de la amígdala (Ame). Posteriormente, el Ame envía señales hacia el hipotálamo anterior (HA) y el núcleo del lecho de la estría terminalis (BNST), para que, a su vez, estos envíen información a la sustancia gris periacueductal (SGPA). El sistema incluye una señal inhibitoria que es enviada desde COF a AMe para regular a este último núcleo. **(Figura 2a)**.

Con base en este modelo del sistema cerebral de la agresión se realizó un modelo funcional primario constituido por nodos con entradas y salidas **(figura 2b)** que especifican las vías que conectan los diferentes núcleos. Dado que se trata de un modelo funcional no se tomaron en cuenta la morfología de las estructuras cerebrales involucradas o sus distancias anatómicas; en cambio, se consideró cardinal el intercambio encauzado de información excitatoria o inhibitoria realizado por las vías aferentes y eferentes de conexión entre dichas estructuras.

En el modelo funcional desarrollado **(figura 3)** se incluyen dos nodos adicionales que representan las señales de activación (Activ) y salida o expresión (Expre) del sistema. Las líneas dobles recortadas que entran y salen de las estructuras internas del circuito reflejan los eventos o señales que ingresan al cerebro y salen de éste para expresarse en los diversos comportamientos agresivos. El conector del nodo COF al nodo AMe simula en la red de Petri la conexión inhibitoria que ocurre según el modelo de Nelson y Trainor [25].

En la **Tabla 1** se comparan las características de las redes de Petri y las condiciones del sistema cerebral definido para la agresión. La primera columna muestra las características básicas de una red de Petri. En la segunda columna se da una breve explicación de la característica asociada de la red Petri para ser acorde con el comportamiento neurobiológico.

## Modelación del sistema cerebral de la agresión mediante una red de Petri

La simulación fue hecha con el software de implementación de redes de Petri (PetriNet Simulator by Biogenerics. Inc.) y los datos fueron analizados con un software especializado (Mathematica 7.0, Wolfram Research).

De acuerdo con la teoría de procesos pautados, los núcleos o módulos cerebrales se representan como los lugares o plazas P de la red pues representan los sitios que generan las activaciones o inhibiciones de cada nodo o estructura cerebral. El conjunto de conexiones sinápticas entre dos estructuras se representa por las transiciones T. Los tokens son las unidades de transición entre nodos y representan paquetes de información de la red neuronal de cada núcleo en forma de una señal específica sea de activación o de inhibición. Para modelar adecuadamente el circuito y contar con suficientes tokens, se planteó una red de capacidad infinita.

De acuerdo con el procesamiento funcional que normalmente ocurre en redes neuronales *in vivo*, la información de tipo concurrente y la retroinformación entre estructuras son también factores necesarios para que funcione el circuito artificial. De esta forma, a continuación se detallan algunas variables funcionales del modelo que se eligieron para emular de manera verosímil el funcionamiento del circuito de la agresión. Además de los módulos básicos que constituyen el circuito, se introdujeron en el modelo dos plazas de input (ACTIV1 y ACTIV2) que representan a las cortezas sensoriales que inician el procesamiento nervioso que desemboca en el comportamiento agresivo. La plaza ACTIV1, por ejemplo, es generatriz de un token en lapsos de tiempos aleatorios, mientras que ACTIV2 es generatriz de cantidades aleatorias de tokens en lapsos de tiempos igualmente aleatorios. El objetivo es emular los ingresos de información en forma de estímulos relevantes que pueden acceder al cerebro en cualquier momento.

La **figura 4** muestra en líneas no punteadas las conexiones que van de los núcleos de activación hacia COF; por lo tanto representan a las vías que convergen en la corteza órbito frontal. Los módulos de entrada ACTIV1 y ACTIV2 envían información a AMe **(Figura 5)**. Además, COF tiene influencia sobre AMe por medio del comparador TTL. Este comparador TTL es un módulo ficticio requerido en el modelo (plazas de recurso), usado para emular la inhibición de COF sobre AMe mediante la comparación binaria de tokens en el tiempo. Es decir, los tokens que modelan la información nerviosa transmitida entre los módulos representados deben primero pasar por la corteza orbito frontal y, dados los valores de las latencias temporales en este nodo, son enviados luego a AMe. Las latencias entre las estructuras COF y AMe generan retrasos y por tanto desfases temporales entre los tokens de los diferentes núcleos para simular las inhibiciones que COF tiene sobre AMe y así poder modelar la concurrencia de la información dentro de la red. En general, los tokens enviados de COF a AMe y que simulan el control de la corteza frontal sobre la amígdala son a su vez controlados por los retardos que tienen lugar en el nodo controlador (TTL).

En general, las características funcionales de los nodos del circuito se basan en datos neurofisiológicos [4, 25], excluyendo los nodos no operacionales (ficticios, pero requeridos para emular el sistema) que sirven como ajustes verosímiles realizados en el modelo: nodos de activaciones, el comparador TTL, las funciones de recaptación (o reverberación) y de expresión.

Entre las características del modelo se incluye una conexión entre la corteza orbitofrontal (COF) y el hipotálamo anterior (HA) que se ha descrito en humanos [4, 18, 25] (**figura 6**). Como se explica a continuación, la plaza TTL es de tipo *recurso* pues está asociada con tiempos de cada una de las estructuras COF y AMe que permiten las latencias entre éstas **(figura 7)**.

El circuito reverberante en este modelo tiene la opción de tomar alguno de los tres valores (**-1, 0, 1**), que muestran los niveles de activación de TTL, la cual es alimentada continuamente por COF y AMe. Estos valores representan un retardo, concurrencia y anticipo de tokens de la siguiente manera: -**1** si un token que viene de OFC tiene un retardo respecto a un token que venga de MeA; **0** para el flujo normal del token en la red a través de ese nodo o para la concurrencia espacio-temporal de tokens y **1** si un token que viene de OFC esta adelantado respecto a un token que venga de MeA.

La plaza de control Recap **(figura 8)** representa el proceso de recaptura de tokens en una parte del circuito que permite alimentar continuamente mecanismos de reverberación. El grafo autocíclico representa la recaptura de tokens en la transición T1 que integra la información que va desde ACTIV2 y TTL hasta AMe. Así, en aras de evitar pérdidas de información, el nodo Recap ayuda a mantener la información en forma de tokens dentro de toda la red y se ubicó en T1 porque a esta transición llega la información de todas las activaciones iniciales.

Luego que la información ha pasado por el circuito reverberante, al llegar a AMe es dirigida a las estructuras LNET y HA **(figura 9)**. Es indispensable que las activaciones de estos núcleos (representadas por los tokens), se den en forma concurrente y en consecuencia se dispara la transición asociada con LNET y HA. De esta manera el proceso de recaptura es usado en el diseño del experimento computacional de forma global para alimentar continuamente el circuito reverberante y poder mantener estable el flujo de información en la red.

La activación de T5 envía tokens a SGPA que, al activarse, genera finalmente la conducta agresiva. Cada una de las plazas tiene tiempos de simulación diferentes de forma que la generación de la conducta agresiva depende de que el envío de información a SGPA sea de forma concurrente y sincrónica, en lapsos de tiempo muy cortos e independientes de los tiempos asociados con cada plaza. Finalmente, la información pasa desde SGPA hasta la representación de la plaza asociada al sistema motor EXPRE **(figura 10)**. Lo anterior se asume bajo la consideración que la expresión de conductas agresivas involucra movimientos voluntarios, pautas de actividad motriz y acciones coordinadas por lo que debe involucrar a sistemas motores piramidales, extrapiramidales y cerebelosos. El módulo EXPRE se identifica con dichos sistemas motores de manera general y no únicamente con la corteza motora, aunque en adelante hagamos referencia solo a esta.

## Criterios operativos de la red

En el algoritmo de simulación los tiempos promedios de respuesta y los tipos de conexiones entre núcleos se tomaron como constantes globales. La entrada del programa permite variar el tiempo total para la simulación en la red. Los datos arrojados o "salidas" del programa son datos numéricos del número de tokens en cada uno de los núcleos neuronales que muestran la relación de la activación en la dinámica de cada uno de los núcleos que conforman la red.

El modelo final requirió de latencias temporales asociadas con algunas plazas y tiempos de latencia relacionados con las activaciones y disparos de COF y AMe en el circuito reverberante, para simular por medio de retardos las inhibiciones de COF sobre AMe y su acción sobre HA. Se hizo una regresión lineal para efectos de comparación entre los tiempos biológicos de transferencia de información y los tiempos de simulación. El promedio ponderado de la pendiente de la regresión generó un factor de escalamiento temporal para obtener la escala de tiempo. De la simulación se extrajeron datos de la activación dada por la cantidad de tokens en un tiempo específico de simulación en cada uno de los núcleos representados por las plazas. Se graficaron estos datos para ver el comportamiento de cada núcleo y para observar el fenómeno de la concurrencia en ejes de activación contra tiempo simulado.

## El factor de escalamiento temporal

Dadas estas características se consideraron velocidades de conducción entre 0.5m/s y 100m/s, lo que conlleva a tiempos entre 2s y 0.01s respectivamente, que es el rango de los tiempos obtenidos en experimentos neurofisiológicos y denominados en este trabajo como rango de tiempos reales de conducción. A su vez, para los tiempos de simulación se usaron intervalos entre 1ms y 30ms asociados con algunas estructuras neuronales, a saber: 30ms para COF, 20ms para AMe, 5ms para HA, y 1ms para LNET. En la simulación estos tiempos se usan para retener y procesar la información correspondiente a los tokens en cada una de estas estructuras. De la misma forma, para la simulación del circuito reverberante se tuvieron en cuenta los tiempos de latencia generados entre las estructuras COF y AMe mediante el comparador, es decir, los valores que determinan cuánto tiempo en ms le toma al comparador TTL dejar de estar activado después de finalizar sus tareas u operaciones. Por ejemplo, al comparador TTL le toma 5 ms inactivarse después de haber estado enviando información a COF y 10 ms inactivarse después de enviar información a AMe. Esto está de acuerdo con el planteamiento inicial de la red, en la cual COF debe estar menos tiempo inactivo que AMe para poder controlar su activación y esto último se refiere al control fisiológico que COF mantiene sobre AMe [25].

La necesidad de correlacionar los tiempos usados en la simulación con el rango de tiempos reales de conducción de información neuronal tiene como base la relación dada por la siguiente expresión:

$$t_s = nt_r$$

donde *n* representa el factor de escalamiento entre los tiempos de simulación ($t_s$) y los tiempos reales de conducción ($t_r$).

A continuación se normalizaron las unidades de los tiempos en ms haciendo estimaciones de la escala lineal entre ellos. La gráfica de la **figura 11** se generó a partir de los datos de los tiempos de simulación vs los tiempos reales de conducción a los que se les hizo una regresión lineal para obtener el factor de escalamiento entre el tiempo de simulación y el tiempo real. Se obtuvo la relación lineal de la siguiente ecuación:

$$t_s = 0.0147938\, t_r + 0.447731 \qquad (ms)$$

y el factor de escalas temporales $n$ con el valor asociado en la siguiente igualdad:

$$n = 0.0147938$$

que representa la tasa a la que aumentan los tiempos de simulación con respecto a los tiempos reales del sistema biológico, es decir, $\Delta t_s \cong 0.01\, \Delta t_r$. La tasa muestra que la variación de los pasos computacionales es aproximadamente una centésima de la variación de los tiempos biológicos reales.

# Resultados

## Dinámica de la red

Las gráficas siguientes corresponden a dos variables definidas como la cantidad de tokens en una estructura específica (estado de activación), representada en el eje vertical y los pasos de la simulación (paso computacional marcado por los "disparos" de la red). La simulación se hizo en un tiempo de 20000 ms a pasos de 20 ms para un total de 1000 repeticiones, -en adelante iteraciones-, en 20 segundos **(figura 12)**.

La gráfica muestra las activaciones de las plazas Activ 1 y Activ 2 que tienen una función generatriz de tokens de tipo aleatorio y periódico, respectivamente. Así, mientras la plaza Activ 1 tenía una función generadora de tokens de tipo aleatorio en las iteraciones, la plaza Activ 2 tenía una función generadora de tipo aleatorio tanto en las iteraciones como en la cantidad de tokens que pudiese generar. De esta forma se representaron los ingresos de información al circuito de la agresión pues, como se sabe, el cerebro está sometido a una infinidad de estímulos que ingresan con un patrón posible al propuesto en la simulación.

Las activaciones de las plazas input envían información conjuntamente a las plazas operacionales que representaban los núcleos COF y AMe los que, dentro de un circuito reverberante, presentan activaciones en términos de las iteraciones **(figura 13)**.

La dinámica de las activaciones en términos de las iteraciones de la simulación muestra una baja probabilidad de activaciones simultáneas en COF y AMe. De hecho, las activaciones de COF producen caídas en las activaciones de AMe lo que genera un comportamiento de tipo pulsátil entre estas dos estructuras. Las activaciones de COF impiden las de AMe, que es el módulo encargado de enviar información a las demás estructuras, pero solo si COF lo permite.

Las activaciones e inhibiciones de AMe mediadas directamente por COF son el resultado de las interacciones de la información dentro del circuito reverberante. Para el diseño del circuito reverberante se tuvo en cuenta el efecto de la recaptura de información, es decir de la información recobrada por reverberación, y el efecto de comparación de información entre núcleos neuronales.

Cabe resaltar que la plaza que representa la recaptación o recobro de información (**figura 14**), y la plaza que representa el comparador TTL no son operacionales pues no son estructuras del circuito de agresión, sino plazas de control y de recurso respectivamente. El comparador TTL mantuvo tres niveles de activación predeterminados donde sus valores de 1, 0, −1 se asocian con niveles de uso. De esta manera, como salida de la simulación y en un rango de 1000 iteraciones en 20 segundos el porcentaje de utilización del comparador fue de 54.2%, lo que indica que la función de comparar información entre módulos es fundamental para generar la concurrencia necesaria para que se logre activar la conducta agresiva.

Retomando la idea funcional del circuito propuesto de la agresión, diversos estímulos externos llegan a las cortezas auditiva o visual enviando información a la corteza orbito frontal y a la amígdala media. Estas estructuras intercambian información mediante un circuito reverberante capaz de comparar información, lo cual genera latencias entre la información que viaja desde COF hasta AMe y HA, inhibiendo a estas estructuras e impidiendo la generación de la conducta agresiva. Así, la clave para que se exprese la conducta de agresión por SGPA depende de la concurrencia de la información en las estructuras LNET y HA, pues no es suficiente la activación de solamente una de ellas **(figura 15)**.

La figura 15 muestra, de arriba hacia abajo las activaciones en términos de los pasos de computo de LNET, HA, SGPA divididas en una magnificación de las mismas en la parte izquierda y su escala normal en la derecha. Es decir: los óvalos que encierran las regiones de las activaciones de la derecha son magnificadas en la izquierda para mostrar la concurrencia espacio-temporal de tokens. El óvalo superior proyecta en la izquierda un evento exacto en el que LNET con cuatro tokens de activación concurre con una activación de un token en HA a los 351 ms. El evento previo de concurrencia en las plazas anteriores permite que la transición T5 se active y dé paso exactamente un milisegundo después a la activación de SGPA con un token. Este token viajará después por medio de la transición T6 al nodo EXPRE, el cual refleja la producción de la conducta agresiva. Se hace énfasis del momento preciso en el que la activación de LNET concurre con la activación de HA, dando paso a la activación de SGPA. De esta forma se pone de manifiesto que solo un proceso pautado debido a envíos de información por estructuras cerebrales jerarquizadas en tiempo discreto, en lapsos cortos y en forma concurrente puede generar la conducta de agresión.

## Discusión

La red de Petri usada para simular el circuito de agresión es una *red temporal discreta de tipo aleatorio-periódico*, dotada además de un grafo autocíclico. Es una red *ordinaria* por tener todos los pesos de sus arcos con un valor de 1 y de *capacidad infinita* debido a que la cantidad de tokens que se pueden almacenar en cada nodo no tiene un límite estipulado.

Aunque el algoritmo usado está de acuerdo con el sistema fisiológico ya que toma en cuenta sus principales características, el papel funcional que tiene cada una de las estructuras del circuito relacionado con la conducta agresiva es incierto ya que ninguna de ellas se involucra sólo en este comportamiento, sino en muchas otras funciones del cerebro. De forma aledaña, la dinámica del circuito correlacionado con la conducta agresiva se puede considerar de tipo *emergente* puesto que el sistema está compuesto de múltiples partes interconectadas en las que la información adicional resultante se oculta al observador. En efecto, la simulación revela un tipo de información que estructuralmente no era evidente. El comportamiento emergente se debe a que a partir de sus interacciones la red muestra una dinámica que no se podría explicar aislando sus partes. La heterogeneidad en las conexiones, que es una de las características de las redes complejas, se evidencia en la presencia de nodos con muchas, pocas y medianas conexiones. El comportamiento emergente requiere que las entradas y salidas de la información en el sistema sean comandadas por funciones lógicas, algunas de las cuales pueden ser descritas y otras solamente se evidencian por el comportamiento de la red entera y no por la causalidad lineal de elementos aislados del sistema. De esta manera, el circuito neurobiológico de agresión modelado en este trabajo es un sistema complejo debido a que está en continuo intercambio de energía o de información con el entorno, posee un comportamiento no lineal e imprevisible y tiene propiedades emergentes. La generación final de la conducta agresiva aparece como resultado de un proceso holístico y jerárquico porque se demuestra que la generación de la expresión de la conducta de agresión es llevada a cabo después de procesamientos de información en los diferentes niveles.

De acuerdo con las evidencias de la neurociencia, este modelo incorpora diversos circuitos de comunicación entre módulos o cúmulos neuronales. Hay circuitos lineales como la conexión entre AMe y el núcleo del lecho de la estría terminal y circuitos convergentes en un módulo debido a la actividad eferente de diferentes núcleos. Así, no se puede generar la conducta de agresión sólo desde la plaza SGPA si no concurren los impulsos de LNET y de HA al mismo tiempo en la dinámica de la red. Se observa, también, un circuito reverberante que permite que la información de un núcleo de neuronas module la información que pasará a un posterior grupo de neuronas al salir de este circuito.

El estado final del sistema modelado surge tanto de las relaciones dentro la red neuronal como de las estructuras involucradas y sus dependencias temporales a partir de estímulos que ingresan del exterior. El modelo también permite validar el efecto de un *macroestado computacional* [19] que se supone subyace a la cognición y a la conducta organizada en función de las relaciones de tiempo y de activaciones en curso de la red neuronal [26]. Los resultados de esta investigación permiten afirmar que la generación de la conducta de agresión evidencia un comportamiento complejo que, simulado mediante redes de Petri, representa una variedad de sistemas dinámicos concurrentes.

El modelo presentado en esta investigación contribuye a validar la teoría de procesos pautados mediante la simulación del circuito nervioso de agresión debido a que su dinámica muestra secuencia de pautas, transformación de información, actividad cinética y periodicidad entre estados nerviosos en secuencia. Según la teoría, los correlatos cognitivos o conductuales del proceso cerebral ocurren por una serie de procesos y eventos discretos y con lapsos de tiempo cortos que se presentan y detectamos como si los eventos fueran continuos [19]. El estudio interdisciplinario de pautas funcionales dinámicas, de su formación, características y evolución en el sistema nervioso desde la perspectiva de la teoría de la complejidad puede llevar un mejor conocimiento de los procesos nerviosos, conductuales y mentales de los que se ocupa la neurociencia cognitiva.

## Bibliografía

**Figuras**

**Figura 1**: Grafo bi-partito en el que se muestran las vías o arcos y las transiciones entre cada plaza o nodo. Los tokens representan unidades dinámicas de intercambio entre las plazas o nodos que comparten una transición y su ubicación define el estado del sistema.

**a.** **b.**

**Figura 2**: **a.** Circuito funcional de estructuras relacionadas con la conducta de la agresión (Modificado de referencia 28). **b.** Primera aproximación al modelo, los signos + y – representan activación e inhibición, respectivamente.

**Figura 3**: Imagen del estado dinámico de la red de Petri que representa el circuito de agresión. Las latencias entre los núcleos se representan por las plazas semi-sombreadas. La cantidad de área sombreada que se va cubriendo en sentido de las manecillas del reloj, representa la duración de la latencia de los nodos al recibir tokens. Una plaza se considera activada con al menos un token en ella como se muestra a manera de ilustración en Recap, LNET y EXPRE.

**Figura 4**: Vías de entrada a COF desde los núcleos de activación Activ 1 y Activ 2.

**Figura 5**: Vías de entrada a AME desde Activ 1 y Activ 2.

**Figura 6**: Vías que van desde COF hacia Ame y HA.

**Figura 7**: Circuito reverberante entre COF, Ame y el comparador TTL representado por las líneas resaltadas.

**Figura 8**: Representación del circuito de recobro como un grafo auto cíclico.

**Figura 9**: Vías que de AMe van hacia LNET y HA.

**Figura 10**: Vías finales del circuito en el caso de concurrencia espacio-temporal de tokens en las plazas LNET y HA. La activación de estas plazas permite que la transición T5 envíe un token a SGPA y por ende a EXPRE.

**Figura 11**: Regresión lineal de los tiempos de simulación vs tiempos reales de conducción de señales. El eje vertical representa el rango de valores de tiempos de computo en los nodos de la red usados en la simulación y el eje horizontal asocia los rango de tiempos de conducción reales.

**Figura 12**: Dinámica de primera y segunda activación, Activ 1 y Activ 2, respectivamente. En esta figura y las siguientes el eje vertical representa la cantidad de tokens que están en cada nodo en un determinado instante de la simulación que dura 20000 ms en pasos de a 20 ms para un total de 1000 iteraciones. Nótese que la actividad en Activ 1 es aleatoria temporalmente y siempre genera un solo token, al tiempo que Activ 2 es aleatoria en el tiempo y genera una cantidad de tokens aleatoriamente.

**Figura 13**: Dinámica de COF (Puntos negros) y AMe (Cuadrados grises) en términos de los tiempos de los pasos de cómputo. Un token es la máxima cantidad alojada en estos nodos. Nótese que la actividad de AMe es mayor en comparación con la de COF y la coincidencia espacio-temporal de la activación de estos nodos no es recurrente.

**Figura 14**: Dinámica del circuito de recobro, el cual muestra que en todo el tiempo de la simulación (variable horizontal) la activación de tokens (variable vertical) es constante, continua y siempre con un token en este nodo, lo que genera la zona gris.

**Figura 15**: Dinámica y detalle del fenómeno de concurrencia en las activaciones de LNET, HA y SGPA (Arriba hacia abajo). La zona izquierda es una magnificación de las activaciones de estos nodos representadas en la derecha. Los óvalos enfatizan el momento exacto de la concurrencia entre nodos de niveles jerarquizados (óvalo grande) y su consecuente efecto de activación de SGPA (óvalo pequeño) para que ocurra el fenómeno de la agresión. Las dos líneas superiores que llegan a la magnificación finalizan en la concurrencia espacio-temporal de LNET y HA, la línea inferior muestra la consecuente activación de SGPA.

## Tablas

**Tabla 1**: Comparación entre las características de las redes de Petri y las condiciones de la red neuronal.

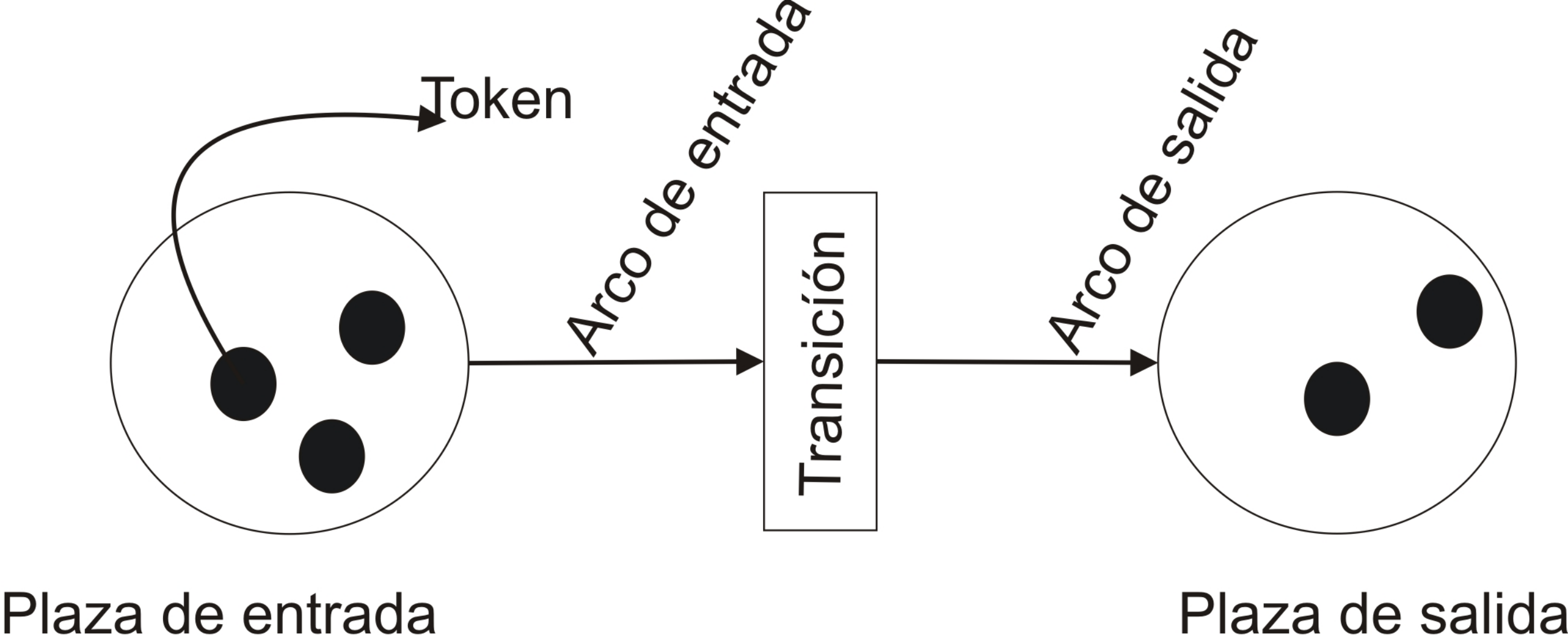

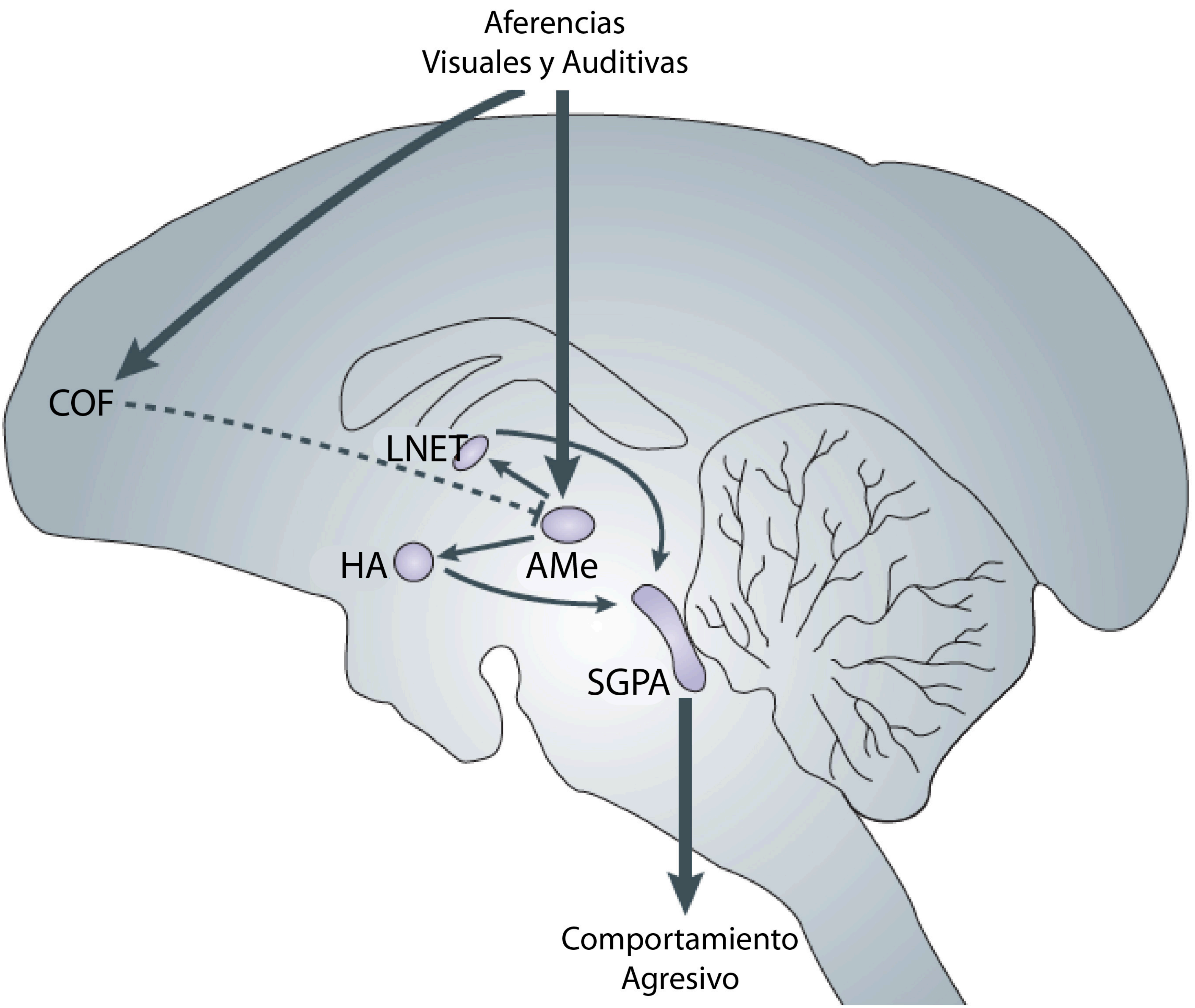

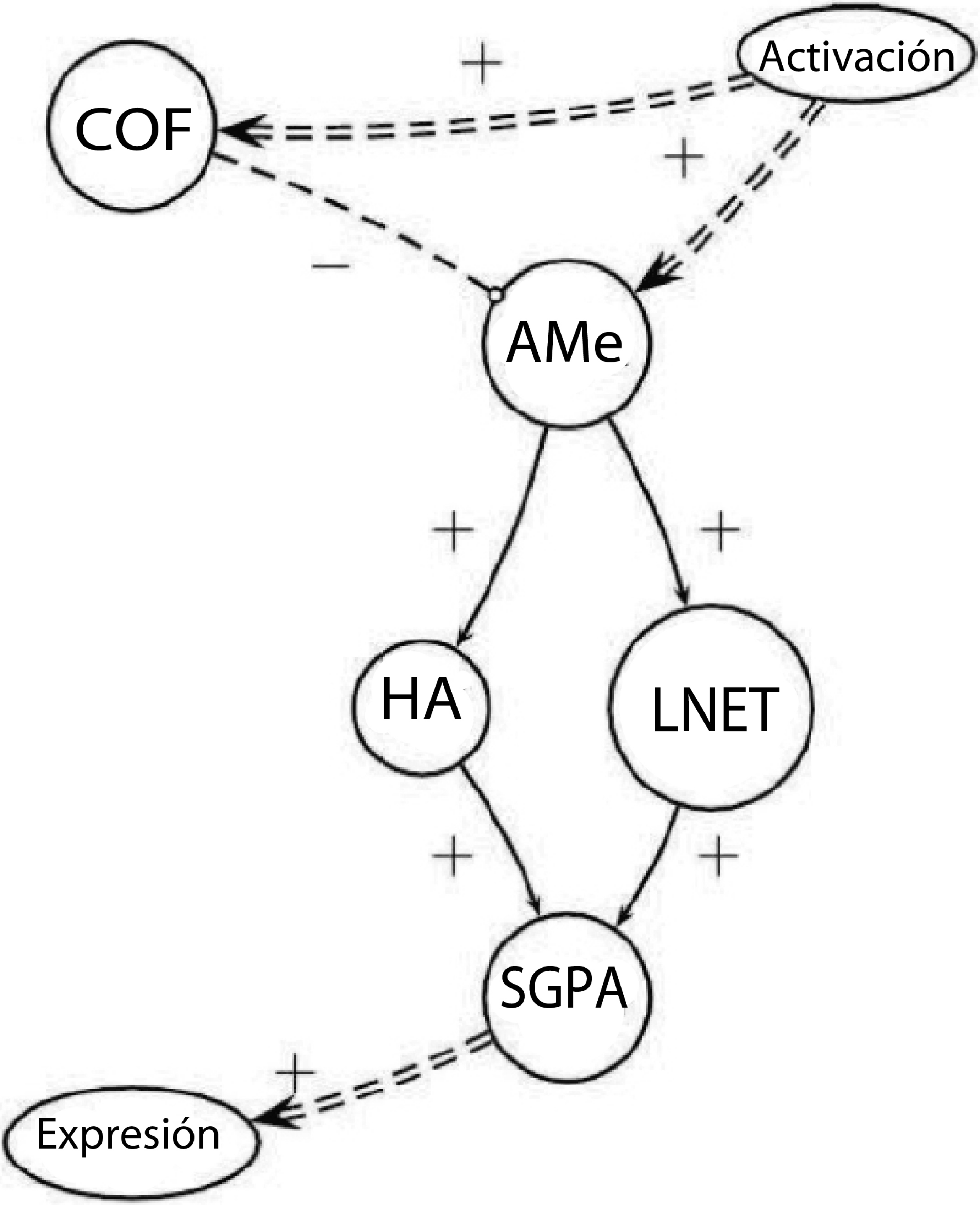

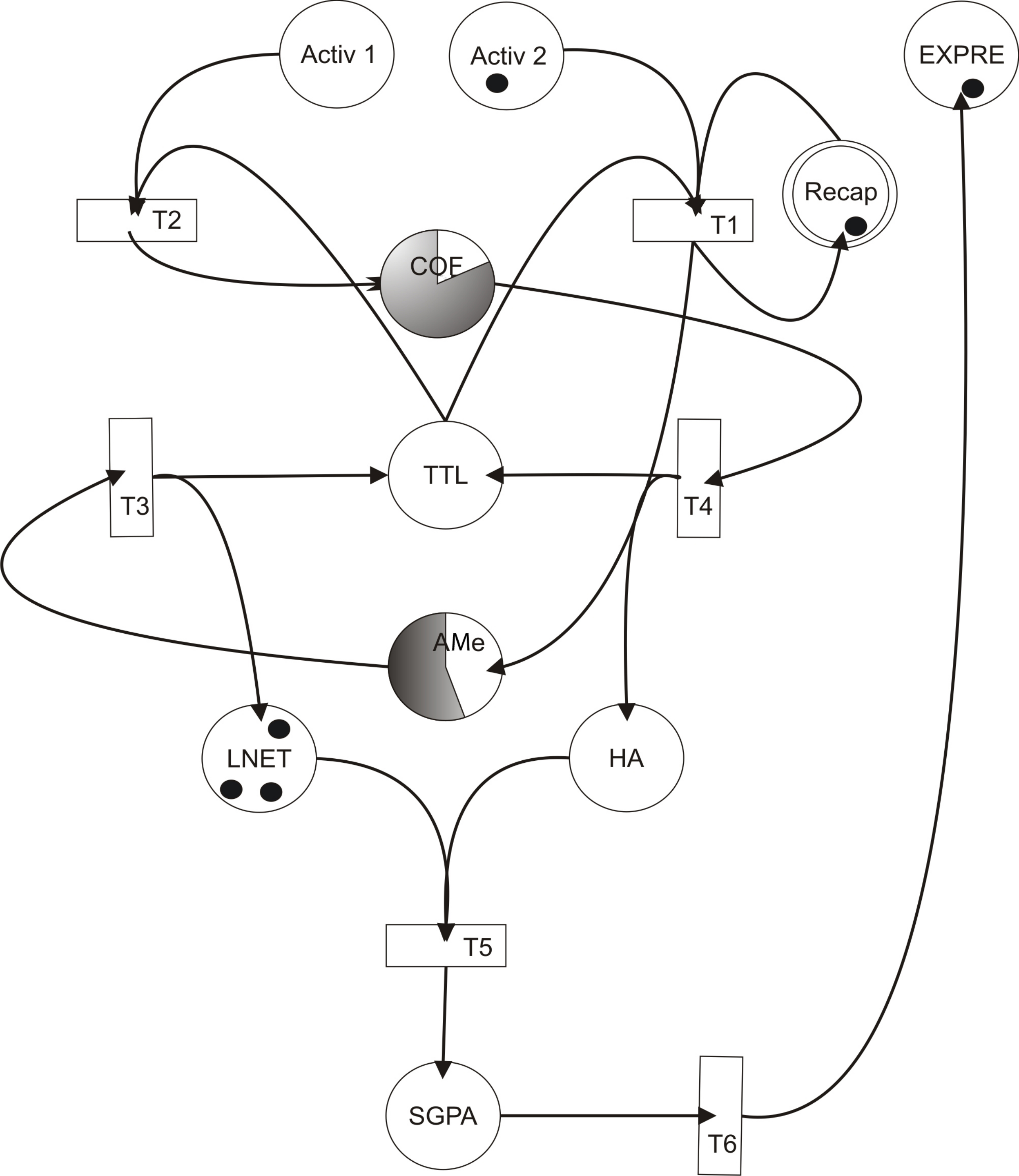

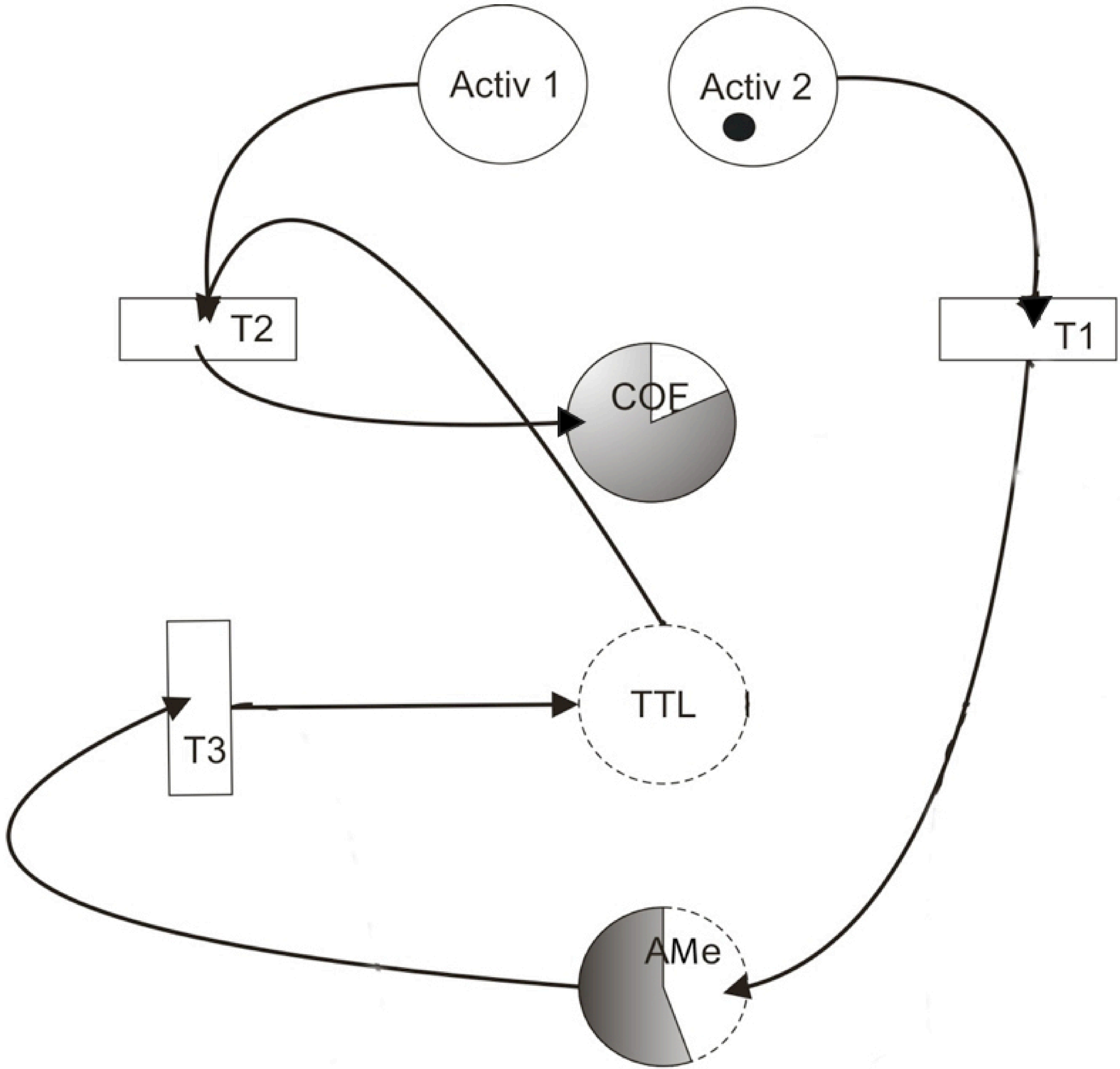

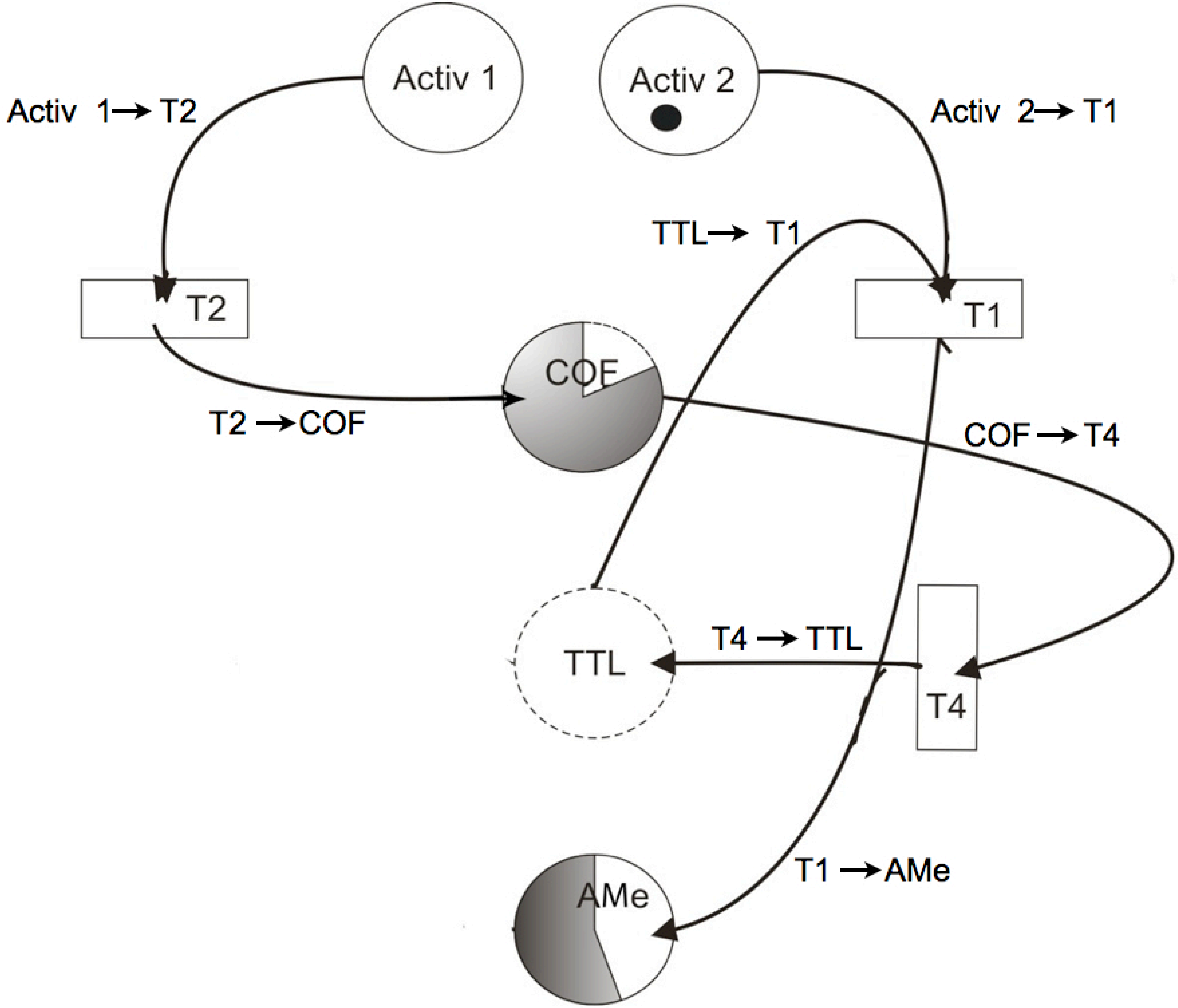

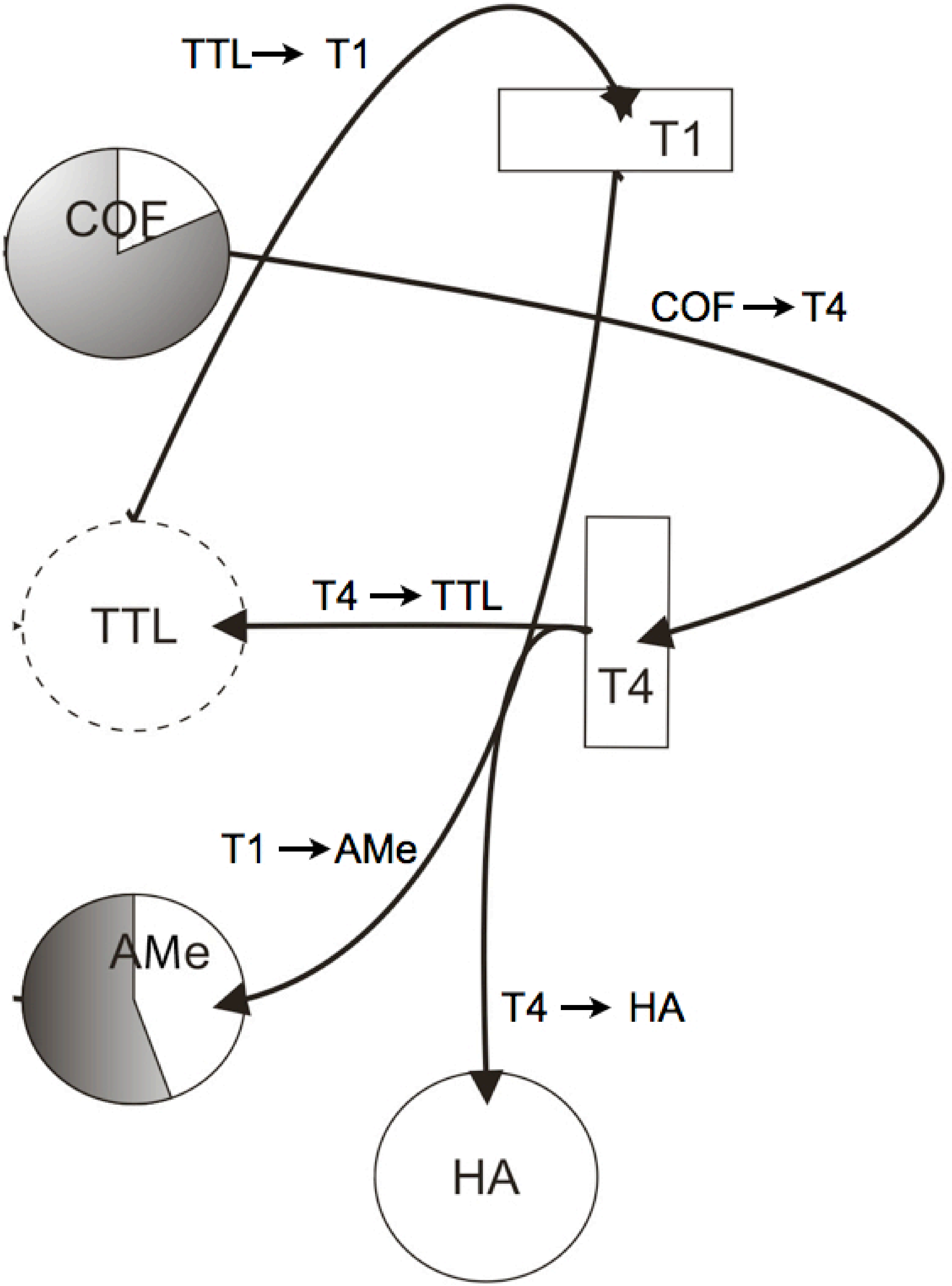

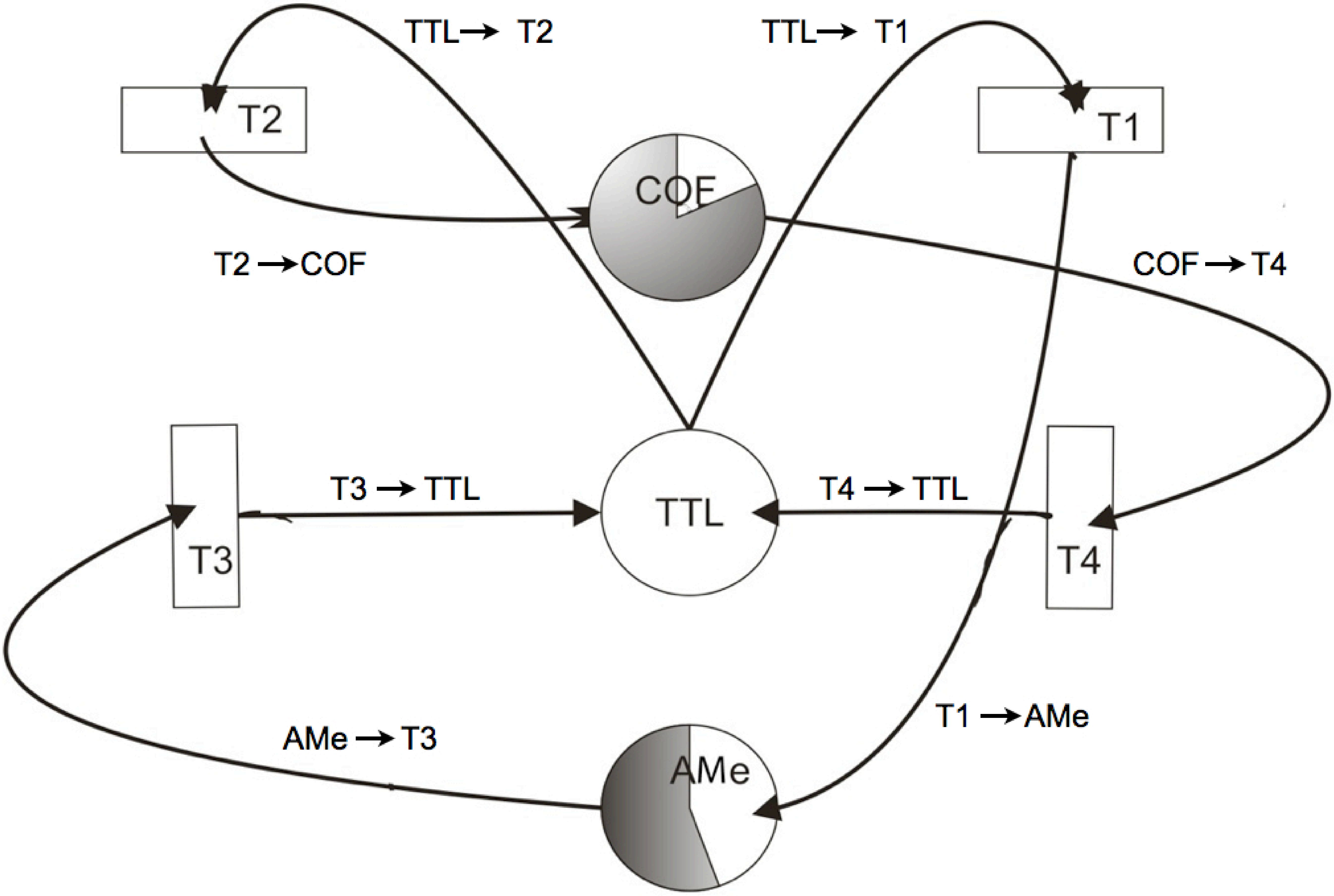

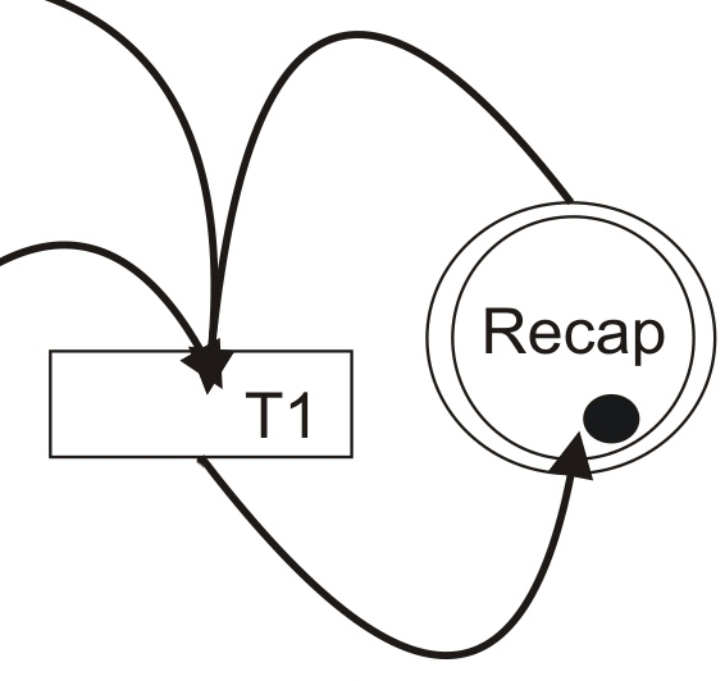

T1
Recap

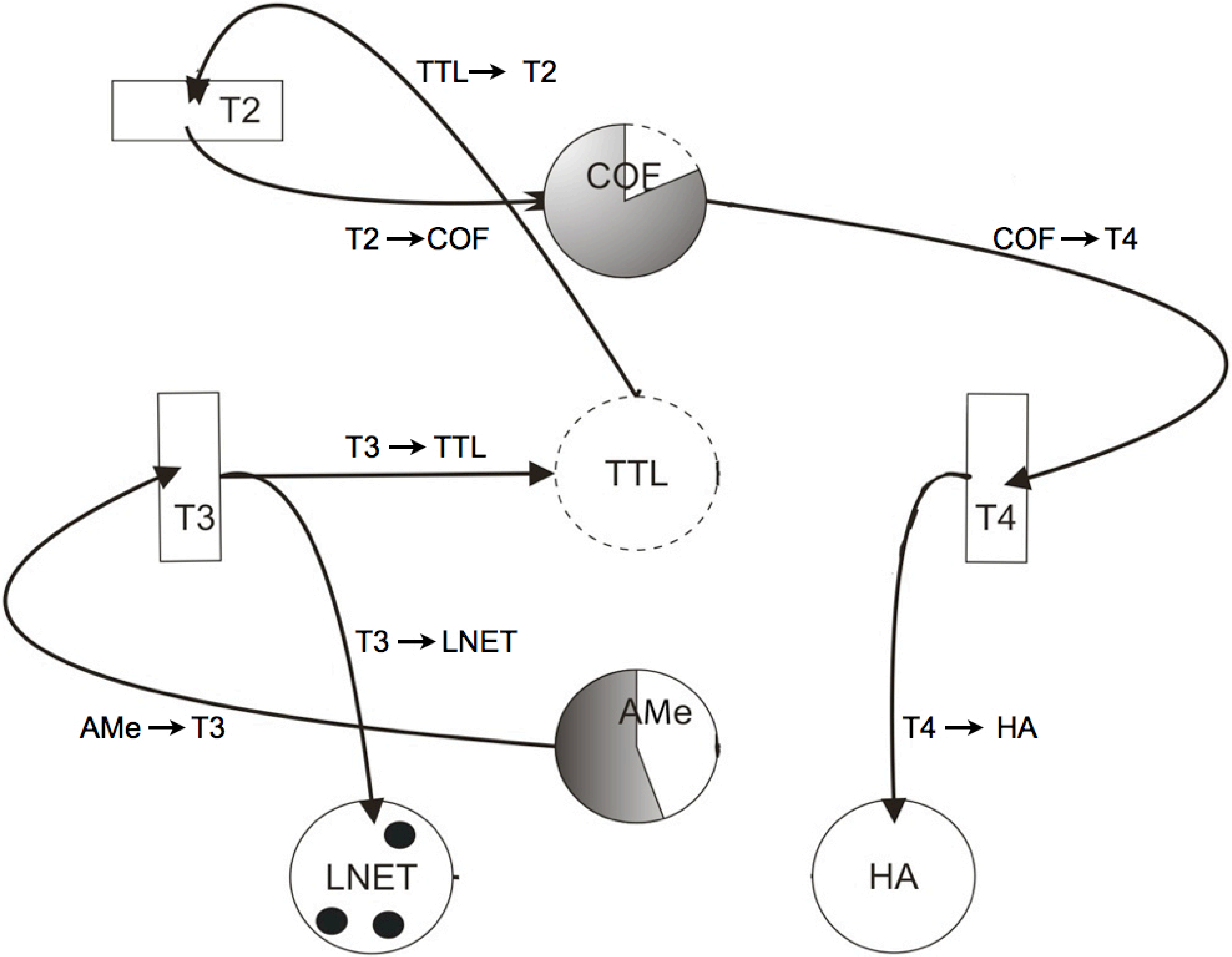

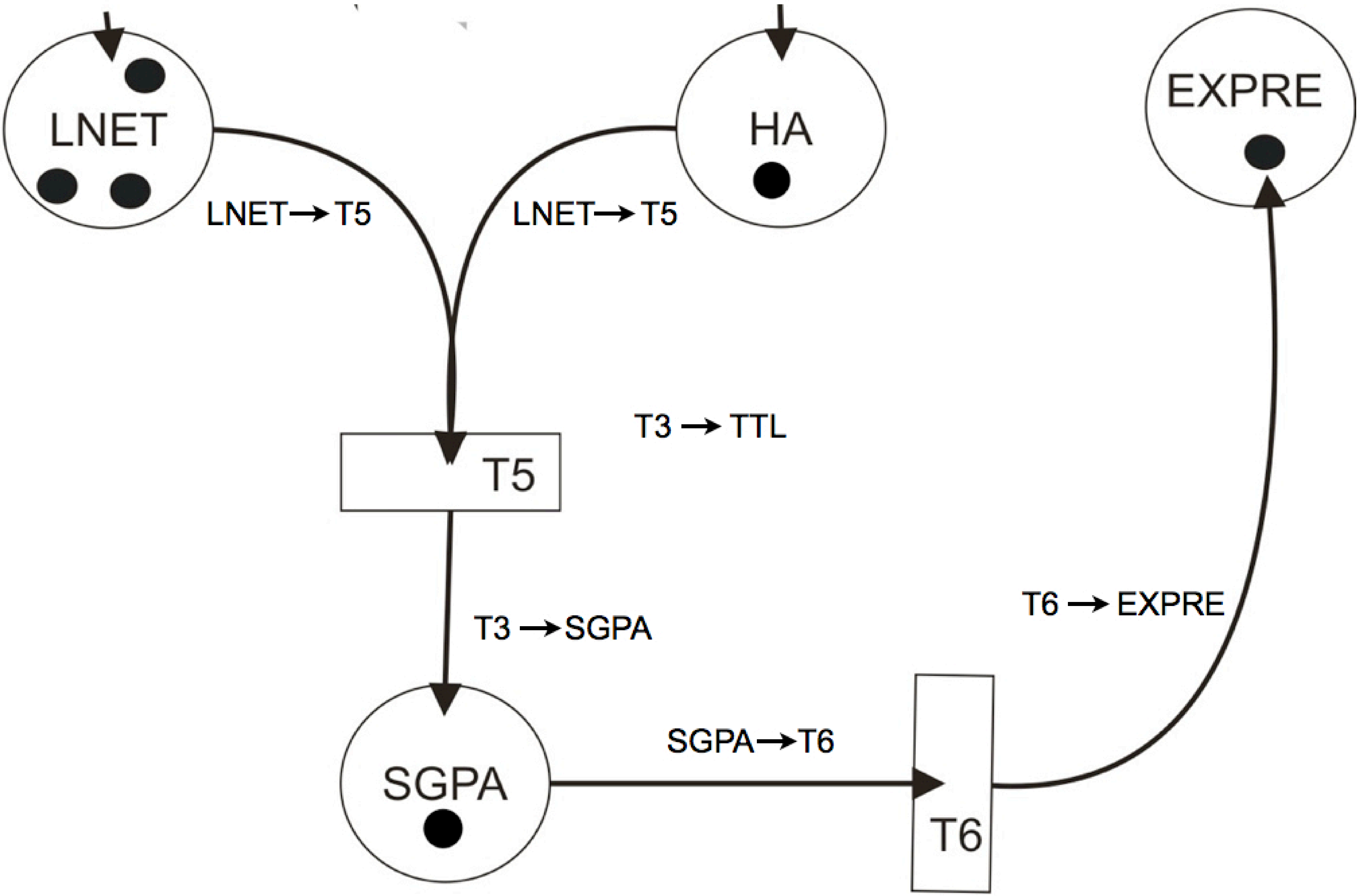

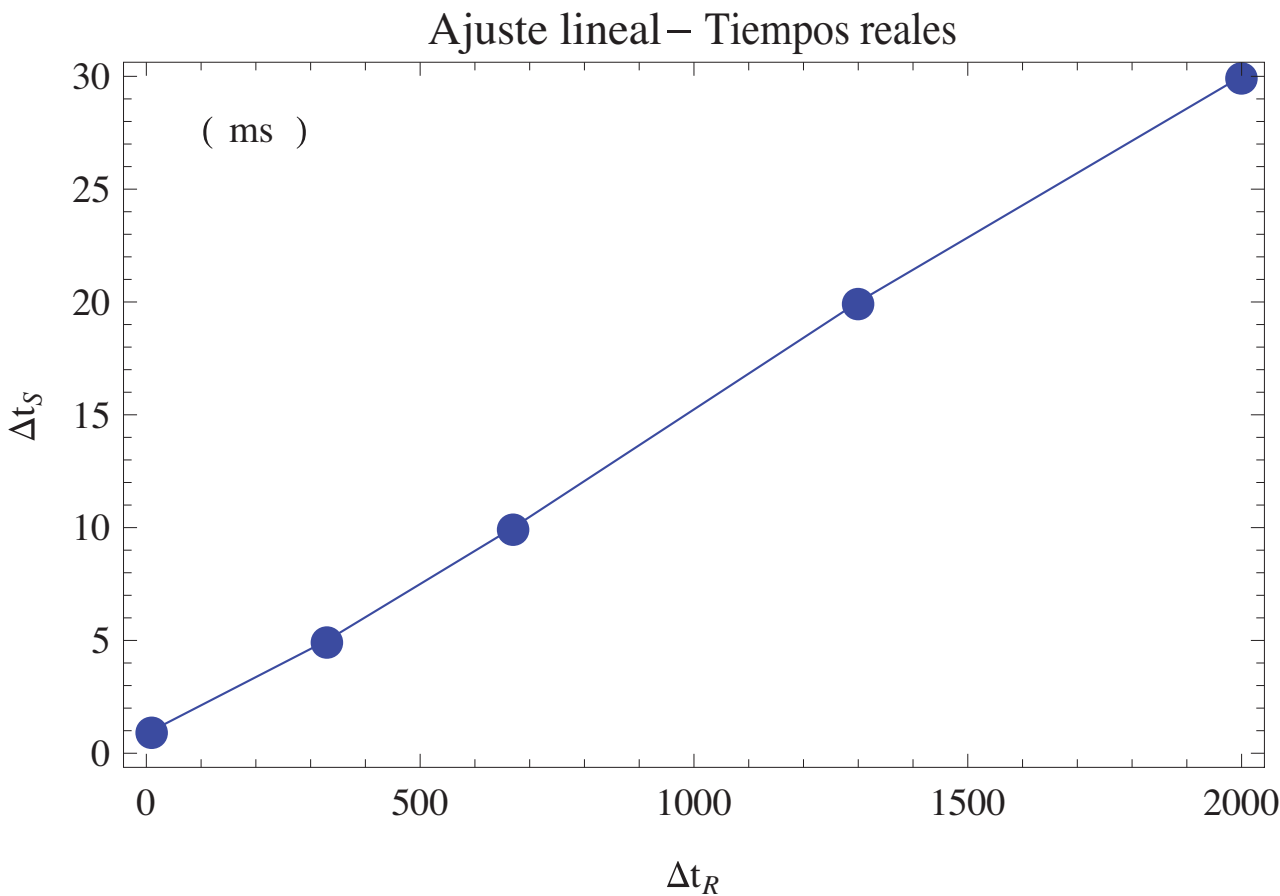

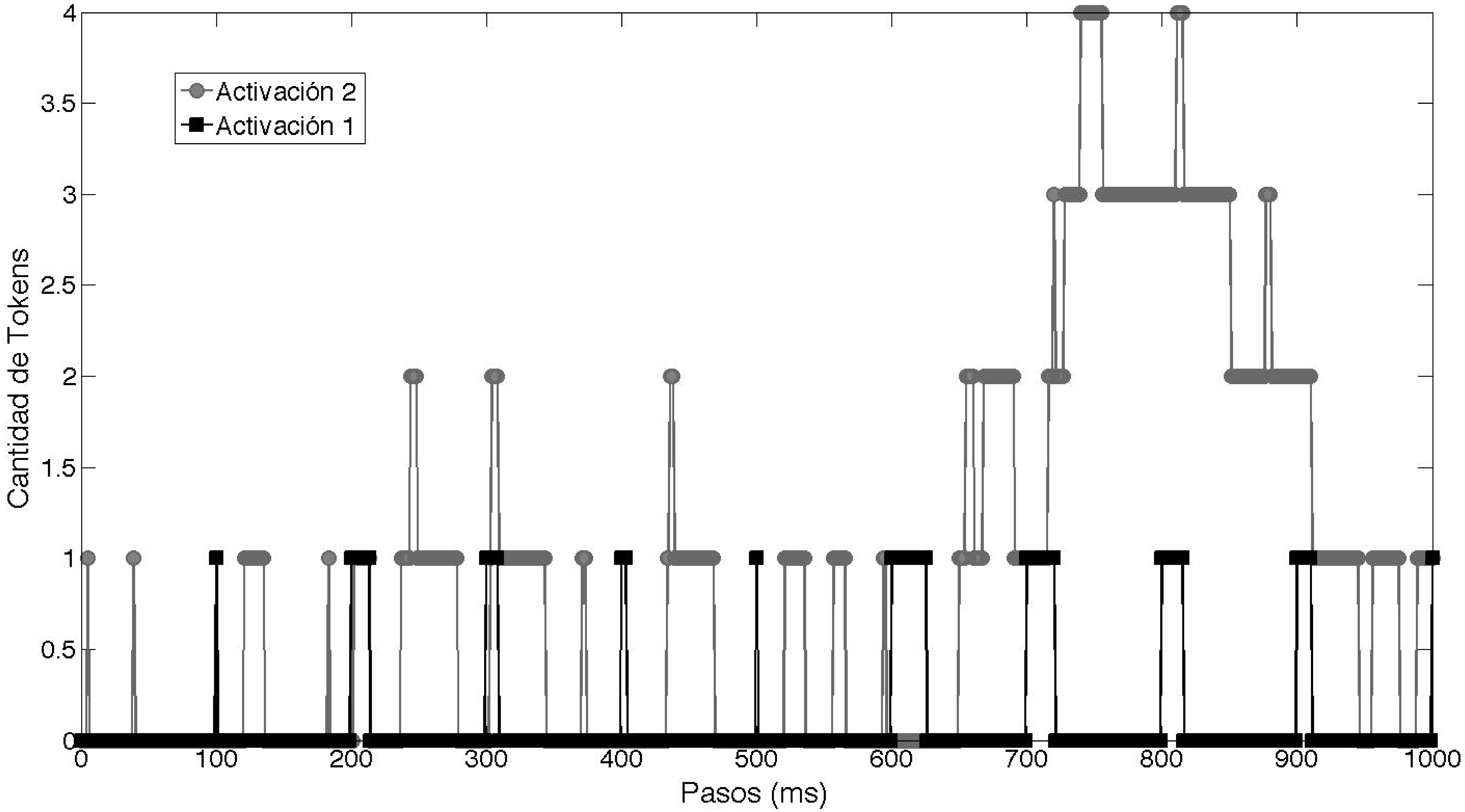

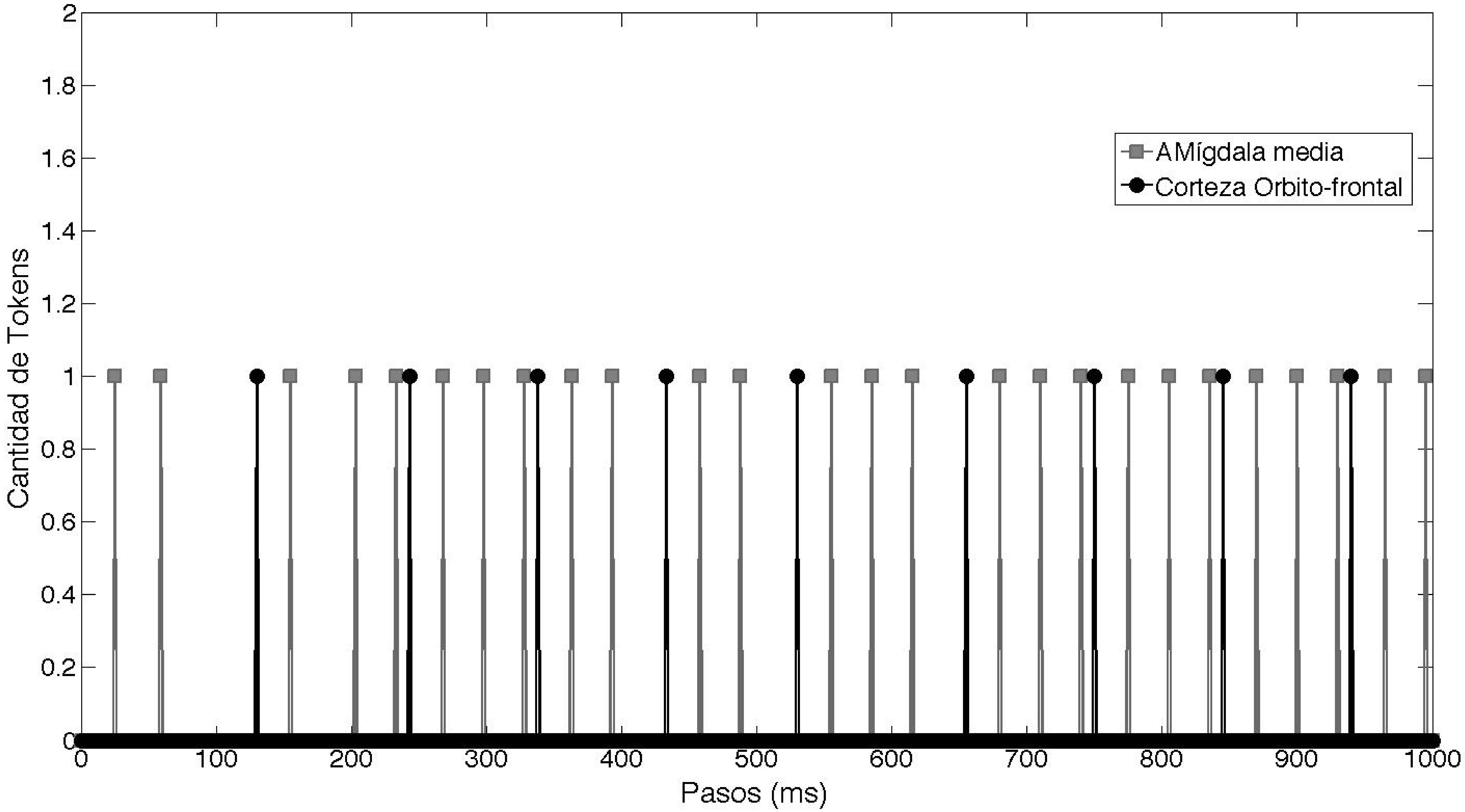

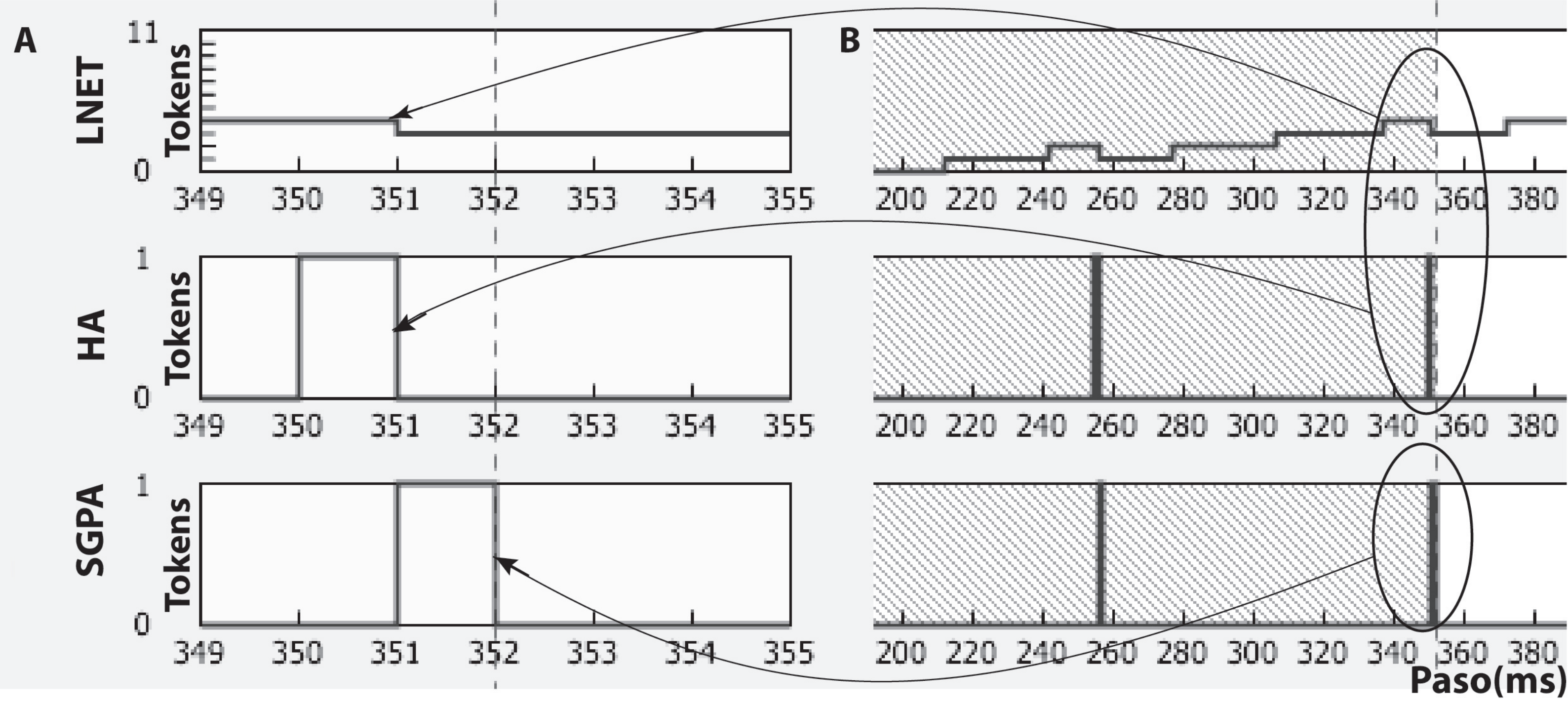

| **Características de red de Petri** | **Características de la aproximación** |
| --- | --- |
| Sincrónico | La expresión es resultado de activaciones, comparaciones e inhibiciones **sincronizadas** de los núcleos. |
| Concurrente | La ejecución de eventos y condiciones es de forma paralela y secuencial (**Concurrente**) |
| Estocástico | La dinámica interna de la red es de tipo no lineal y con aleatoriedad (**Estocástico**) |
| Discreto | La dinámica de la red es **discreta** acorde con los procesos pautados en tiempos infinitesimales |